\documentstyle[preprint,aps]{revtex}
\begin{document}
\def\be{\begin{equation}}
\def\ee{\end{equation}}
\def\tc{{T_c}}
\def\214{{La_2 Cu 0_4}}

\begin{title}
{Resonant Pair Tunneling and\\
Singular effects of c-axis disorder in\
cuprate and organic superconductors}
\end{title}

\author{ G. Baskaran }

\address {The Institute of Mathematical Sciences\\
Madras 600 113, India.}

\maketitle

\begin{abstract}
In the interlayer pair tunneling (ILPT) theory of superconductivity
the large scale  $T_c$ has its origin in the k-space locality of the
inter layer pair tunneling matrix elements.
We reinterpret the same physics as a process of resonant pair tunneling
and illustrate it through cooper pair analysis.
This interpretation is used to give a mechanism which leads to a
singular suppression of $T_c$  as  function of c-axis(off plane/axis)
disorder.  In this mechanism the non resonant tunneling processes
arising from the c-axis disorder in general contributes a pair binding
energy which is reduced by a factor $\frac{T_J}{\epsilon_F}$, where
$T_J$ is the interlayer pair tunneling matrix element and $\epsilon_F$
is the fermi energy. This leads to a simple theorem which states that
the scale of $T_c$ is controlled by the space average value of the
bare one electron interlayer hoping matrix element.  After briefly
discussing  that the ET and TMTSF molecule based organic superconductors are
strongly correlated narrow band systems, the dramatic reduction of
$T_c$ by anion disorder in organic superconductors is explained by our
mechanism.  Off plane disorder effects in some of the
cuprates are also discussed.

\end{abstract}

\section{Introduction}
One of the important component of the RVB theory for the
cuprates is the inter layer pair tunneling (ILPT) mechanism for
superconductivity proposed by Wheatley, Hsu and Anderson\cite{wha} (WHA).
This mechanism owes its existence to the anomalous normal state of the
CuO$_2$ layers which i) suppresses the coherent one electron tunneling at
low energies between two adjacent $CuO_2$ planes and ii) does not
suppress the second order process of coherent pair tunneling. The
blocking of coherent one electron tunneling has been called\cite{conf}
`confinement'.
Anderson suggested\cite{1kgap} sometimes back a BCS like formalism that
incorporates
the physics of interlayer pair tunneling. He also traced the origin of
the large scale of $T_c$ to a k-space local character of the pair
tunneling matrix elements.  Following this the WHA theory got
a recent revival and important applications\cite{sudip} have been made
to the cuprate superconductors.  ILPT processes as a source of
pairing has been invoked in the past for quasi 1-d
organic conductors\cite{iptorg}.

{}From experimental point of view, the anomalous c-axis transport
$\rho_c(T)$ and $\sigma_c(w)$ for cuprates exhibit striking features
suggesting confinement\cite{confexpt}.  Families of quasi 1-d and 2-d
organic conductors, where strong correlation and narrow band character is
manifest, has also been suggested\cite{clark} to exhibit confinement by an
analysis
of the commensurability effects in the angle dependent magneto resistance
in $(TMTSF)_2 X$ family.

The aim of the present paper is two fold:
i) to give a reinterpretation of the origin of large scale of
$T_c$ as a consequence of ``pair tunneling resonance" arising from the
k-space locality of pair tunneling process and
ii) To provide a simple mechanism of how c-axis (off plane or off axis)
disorder can
remove the pair tunneling resonance and lead to a strongly reduced
scale of $T_c$; and
discuss existing experimental results in cuprate  and and  organic conductors
from
the point of view of our new mechanism.
At the end we discuss our result in the light of Anderson's
theorem\cite{pwads} on
dirty superconductors.  We find that the simple cooper pair
analysis, to which we restrict ourselves in this paper, already brings out the
consequences of resonant pair tunneling and and also shows how the off
plane disorder can affect $T_c$ in a singular way. We state our result
in the form of a theorem. We also feel that our explanation of the
anomalous suppression of $T_c$ by off plane disorder gives support to a
substantial contribution from the ILPT mechanism of for superconductivity in
organic superconductors\cite{iptorg}.

\section{Resonant Pair Tunneling and the large scale of $T_c$}

As mentioned in the introduction, the spin-charge decoupled
anomalous normal state prevents coherent one electron tunneling at the
lowest energies. This blocking has been explained as an orthogonality
catastrophy\cite{conf} arising in the two non fermi liquids planes after
the event of
 one electron transfer between the them.  What is remarkable is the
suggestion that this
orthogonality catastrophy is absent when two electrons with zero centre of
mass momentum in a spin singlet state is transferred from layer to
layer in a second order quantum mechanical process.  The presence of
coherent interlayer pair tunneling and absence
of coherent interlayer one electron tunneling is the origin of the novel
WHA mechanism.  Anderson\cite{1kgap} incorporates the key features of the above
physics in a BCS type reduced Hamiltonian.
\begin{eqnarray}
     H  & = & \sum (\epsilon_k - \mu)(c^{\dagger}_{k \sigma} c_{k \sigma} +
d^{\dagger}_{k \sigma} d_{k
\sigma}) + \sum T_J (k) (c^{\dagger}_{k \uparrow} c^{\dagger}_{-k \downarrow}
d_{-k\downarrow} d_{k\uparrow} + h.c. ) \nonumber \\
        &   & + \sum V_{kk'} (c^{\dagger}_{k \uparrow} c^{\dagger}_{-k
\downarrow} c_{-k' \downarrow} c_{k' \uparrow} + d^{\dagger}_{k \uparrow}
d^{\dagger}_{-k \downarrow} d_{-k' \downarrow} d_{k' \uparrow})
\end{eqnarray}
Here c's and d's are the electron operators of the two layers and $k =
(k_x,k_y)$ is the in plane momentum of the electron.
$T_J(k) \approx \frac{t^2_{\perp}(k)}{t}$, is the interlayer pair
tunneling matrix element.  Here $t_{\perp}(k)$ is the inter layer one
electron bare hoping matrix element and $t$ is the in-plane hoping matrix
element.  And $V_{kk'}$ is the residual in plane pair scattering matrix
element which summarizes formally the effect of phonon mediated
and residual correlation induced attraction processes.  For convenience
we will concentrate on two coupled layers throughout this paper.  The
two layer case captures most of the important aspects of an n-layer
system.

The entire physics of spin-charge decoupling, confinement and pair
tunneling is approximately modeled through the presence of pair
tunneling and absence of one electron tunneling terms between planes in
an otherwise fermi liquid like BCS Hamiltonian.  Anderson\cite{1kgap}
argues that
this fermi liquid approximation is a reasonable one below $T_c$ in view
of the fact that the electron propagator changes its branch point
singularity into BCS quasi particle poles.  Recovery of a pole structure
of the propagator is argued to be a self consistent justification for starting
with a fermi liquid like picture to study the superconducting state.

Another important
aspect of the above Hamiltonian is the individual electron momentum
conserving nature of the pair tunneling terms, which Anderson calls as
k-space locality.  This k-space locality, however, does not simply follow from
the non fermi liquid or spin - charge decoupled character of the normal
state of $CuO_2$ planes.  Recently I have argued\cite{gb1k}
that it arises if one
assumes a tomographic Luttinger liquid normal state.

Anderson argued that it is the k-space locality that leads to a scale of
$T_c$ which is linear in the pair tunneling matrix element.  Anderson,
on solving the resulting gap equation in the limit of interlayer
pair tunneling matrix element $T_J$ large compared to $V_{kk'}$, finds
\be
k_BT_c \approx \frac{T_J}{4}~~ for ~~T_J > V_{kk'}
\ee
In the other limit he finds the usual BCS expression
\be
k_BT_c \approx \hbar \omega_D e^{-\frac{1}{\rho_0 V_0}}
\ee
where $\omega_D$ is the Debye frequency, $\rho_0$ is the density of
states at the fermi energy and $V_0$ is the fermi surface averaged
matrix element $V_{kk'}$ of equation 1.

The above result of Anderson for the case $T_J > V_0$ lends itself to
an interpretation of a binding energy arising from splitting of two
degenerate pair states (one from each layer), which are resonantly
coupled by the pair tunneling matrix element $T_J$.  Our interpretation
of $T_c$ enhancement or the corresponding pair condensation energy
or the gap, as arising out of a resonant phenomenon brings out the
singular
effects of off plane disorder in a natural way as we will see in the
following sections.  What we are going to do is a cooper pair analysis
for various cases.  It turns out that this simple minded analysis brings
out the basic physics that we are after.

Mention should be made of phase fluctuations associated with the
ultralocal interlayer pair tunneling processes of equation 1.
This process, as explained
recently\cite{gb1k}, has a local U(1) invariance in k-space and by
itself is not
capable of generating a finite $T_c$.  It needs the help of additional
k-space non local terms such as the third term of equation 1.
What is important is that even with a little help from these non local
terms, the local term can become important
and even provide a large scale for $k_BT_c \approx {T_J\over t}$.
In other words, Anderson's mean field analysis of the Hamiltonian of
equation 1 is meaningful only in the presence of the third term which
is non local in k-space.
Our Cooper pair analysis should be also understood in the above light.
The ultra locality in k-space is only an idealization.  In
general the k-space locality will be smeared by finite
temperature effects or the locality can change from a delta
function type to a power law one\cite{gb1k}

When the two adjacent planes of a bilayer are identical, the k-space
locality of pair tunneling also implies a resonant tunneling of a pair
of electrons between two states $(k,-k) $ and $ (k,-k)$ of the two planes.  To
understand the resonant tunneling, let us consider the Cooper pair
problem with the Hamiltonian given by equation (1).  We consider two
electrons in an otherwise frozen fermi sea of the two layers.  The
Schrodinger equation for the Cooper pair problem is
\be
2(\epsilon_k - \mu) \phi_k - T_J \eta_k + \sum_{k'} V_{kk'} \phi_{k'} = E
\phi_k
\ee
\be
2(\epsilon_k - \mu) \eta_k - T_J \phi_k + \sum_{k'} V_{kk'} \eta_{k'} = E
\eta_k
\ee
where $\phi_k$ and $\eta_k$ are the pair amplitudes in layers 1 and 2
respectively.  We will assume a simple BCS kind of model potential for
$V_{kk'}$: a value $-V_0$ for $k,k'$ lying in an energy shell of $\hbar
\omega_D$ around the fermi surface and zero otherwise.

The above Schrodinger equation is easily solved for
$T_J > V_0$ to get an expression for Cooper pair binding energy:
\be
E_B \approx T_J + \hbar \omega_d e^{-{1\over{\rho_0 V_0}}}
\ee
where $\rho_0$ is the density of states at the fermi level.The first
term is the binding arising from pair tunnel splitting and the second
term arises from the usual BCS type in plane pair scattering processes.
This interpretation is also obvious if we look at Anderson's analysis of
the gap equation and $T_c$ for the above limit.  It is
interesting to note that our explanation of resonant pair
tunneling also brings out the `kinetic' or interlayer
delocalization origin of the pair binding energy of the Cooper
pairs.

\section{Non Resonant pair tunneling and reduction in the cooper pair
binding energy}

The resonant cooper pair tunneling together with the presence of other
non-local terms can lead to a superconducting state with a large $T_c$.
In this paper we will concentrate on how this resonant character of
pair tunneling can be offset by off axis or off plane disorder.
Application to cuprates and organic conductors will be discussed in the
next section.

We model the c-axis or off plane randomness by a position dependent
one electron interplane hoping matrix element $t_{\perp ij}$.
For simplicity we assume $t_{\perp ij} \approx \delta_{ij}
t_{\perp i}$, a short ranged form.  Here i and j are the site
indices of the two planes respectively. It is
important to introduce randomness at the level of bare one electron inter
layer hoping and and see how the pair
tunneling terms that are generated by the physics of the non-fermi
liquid state of the planes get modified.  The bare one electron tunneling
term is

\begin{eqnarray}
H_{12}  &  =  & \sum t_{\perp i} (c^\dagger_{i\sigma}
		 d_{i\sigma} + h.c.) \nonumber \\
        &  =  & \sum t_{\perp}(k,k') (c^\dagger_{k\sigma}d_{k'\sigma} + h.c.)
\end{eqnarray}

The c-axis disorder does not conserve the in plane momentum in the one
electron interlayer  hoping process.
This term. when small compared to the in plane t, which is the case with
the anisotropic conductors under study, does not directly affect the anomalous
normal state of the plane and the nature of the quasi particles.
As we will mention in the last section, this assumption is not really
valid if we have a weakly coupled fermi liquid at zero temperature.
In view of this, while constructing the pair tunneling Hamiltonian we
need not go to scattering eigen state representation, i.e. the
eigen states of the one electron hamiltonian of the two layers
including the random one electron inter layer hoping terms.  At the
level of approximating the in plane physics by a fermi liquid physics
a la' Anderson, the
relevant one electron eigen states continue to be  plane waves.
This is an important difference, when we contrast it with the
situation of Anderson's theorem\cite{pwads} for disordered superconductors.

Using a procedure recently suggested by the author\cite{gb1k},
we get an expression for the pair tunneling term
\be
\sum{t^2_{\perp}(k,k')\over t}
c^\dagger_{k\uparrow}c^\dagger_{-k\downarrow}d_{k'\downarrow}d_{k'\uparrow}
+ h.c.
\ee
The pair tunneling term, while it conserves the in plane center of mass
momentum does not conserve the individual electron's in plane momentum.
Thus the pair tunneling term looses the local U(l) invariance in
k-space.  While this is good for stabilizing the phase fluctuations,
it is not so good in the sense of loosing resonant pair tunneling processes
at the expense of introducing non-resonant tunneling processes
as we will see below.

A general matrix element $t_{\perp}(k,k')$ of equation 8
represents a pair tunneling between two states $(k,-k')$ and
$(k',-k')$ of layers 1 and 2. Since in general $\epsilon_k \neq
\epsilon_{k'}$, it causes non resonant pair tunneling processes.
The reduction in cooper pair binding can be easily estimated by
concentrating on  the pair subspace $(k,-k)$ and $(k',-k')$ of layers 1
and 2.  This is a good approximation  in the limit $T_J(k,k') \gg V_{kk'}$.
The corresponding $2 \times 2 $matrix to be diagonalized is:

\[ \left( \begin{array}{cc}
	    2(\epsilon_k - \mu)  &  T_J(k,k') \\
		 T_J(k,k')       &  2(\epsilon_{k'} - \mu)
		 \end{array}
           \right)  \]
The lowest eigen value of this matrix gives us the new energy eigen
value of a pair
of electron taking into account the tunneling.  The shift in
the lowest eigen value, which is a measure of inter plane pair delocalization
energy  or pair binding energy $E_B$ is given by
\be
E_B  =  \sqrt{ (\epsilon_k - \epsilon_{k'})^2 + T_J^2(k,k')} -
|\epsilon_k - \epsilon_{k'}|
\ee
This pair binding energy is the largest in the resonant case,
$\epsilon_k = \epsilon_k'$ :
\be
E_B = T_J(k,k)
\ee
When it is maximally nonresonant, $\epsilon_k - \epsilon_k' \gg T_J(k,k')$
the pair binding is given by
\be
E_B \approx  \frac{T^2_J(k,k')}{|\epsilon_K - \epsilon_{k'}| }
\ee

We will discuss two simple cases of randomness and calculate
reduction in average cooper pair binding energy:
a) the interlayer hoping parameter $t_{\perp i}$ becomes an uncorrelated
random variable with a mean $<t_{\perp i}> $ and spread $\delta t_{\perp}$
and b) $t_{\perp i} = \pm t_{\perp 0}$ with probability p and (1-p).
And the average $<..>$ symbol stands for a spatial average over the
plane.
For case a,
\begin{eqnarray}
H_{12}  &  =  & \sum  t_{\perp i} (c^\dagger_{i\sigma} d_{i\sigma} +
h.c.) \nonumber \\
        &  =  & \sum  <t_{\perp i}> (c^\dagger_{i\sigma} d_{i\sigma} + h.c.)
       + \sum  \delta t_{\perp i} (c^\dagger_{i\sigma} d_{i\sigma} + h.c.)
\end{eqnarray}
where $\delta t_{\perp i}$ is an uncorrelated random variable with mean
0 and spread $\delta t_{\perp}$.
In momentum representation
\begin{eqnarray}
H_{12} &  =  & <t_{\perp i}> \sum (c^\dagger_{k\sigma} d_{k\sigma} + h.c.)
       + \sum  \delta t_{\perp}(k,k')
       (c^\dagger_{k\sigma} d_{k'\sigma} + h.c.) \nonumber \\
       &  =  & \sum (<t_{\perp i}> \delta_{kk'} + \delta t_{perp}(k,k'))
       (c^{\dagger}_{k\sigma}d_{k'\sigma} + h.c. )
\end{eqnarray}
The first term, the average term, conserves the in plane momentum in the
hoping process.  The second term, the fluctuation term, does not conserve
the in plane momentum.  The one electron tunneling term leads to a pair
tunneling matrix element of the form
\be
T_J(k,k') \approx
\delta_{k,k'}\frac{<t_{\perp i}>^2}{t} +
\frac{(\delta t_{\perp}(k,k'))^2}{t}
\ee
Since we have a spatially  uncorrelated random variable, a
typical value of of $|k - k'| \approx \frac{\pi}{a}$.  For this type of
momentum transfer, the typical value of $\epsilon_k - \epsilon_{k'}
\approx \epsilon_F$, of the order of the fermi energy or band width.
Thus for a typical value of $k \mbox{and} k'$ the pair delocalization energy
is
\be
E_B \approx  \frac{T^2_J(k,k')}{|\epsilon_K - \epsilon_{k'}| }
\approx \frac{T^2_J(k,k')}{\epsilon_F}
= \left( {\frac{T_J(k,k')}{\epsilon_F}} \right){T_J(k,k')}
\ee
The pair binding energy is thus reduced from a resonant value of the order
of $T_J$ to a fraction $\frac{T_J}{\epsilon_F}$ of $T_J$.  For the case
of the cuprate superconductors for a bi-layer of 123 material,
$T_J \sim 40 ~ \mbox{meV}$ and assuming a fermi energy of 2 eV we get
$\frac{T_J}{\epsilon_F} \sim  {1\over{50}}$.  Thus a typical pair binding
energy due to the process of pair tunneling is reduced by a factor of
50.

The first term of equation (14)leads to resonant tunnelling, whereas the
second term is non resonant.  By the fact that the mean $t_{\perp i}$
gets reduced as we
introduce randomness and also the fact that the second term is non
resonant, the cooper pair binding energy and the corresponding $T_c$
decreases.  Since the typical momentum transfer due to the c axis
randomness is large $\approx \frac{\pi}{a}$, the non resonant term
practically leads to no pair tunneling binding energy.  Hence in the
first approximation the pair tunneling binding energy is controlled by
the diagonal value $T_J(k,k) \approx \frac{<t_{\perp i}>^2}{t}$,
the resonant
part of the pair tunneling processes.

For the second type of disorder, arguing in a similar fashion, $T_c$ is
given by
\be
k_B T_c \approx \frac{<t_{\perp i}>^2}{4t} =
\frac{(p -\frac{1}{2})(t_{\perp 0})^2}{4t}
\ee
We summarize the discussion of this chapter in the form of a
theorem: `The transition temperature in the interlayer pair
tunneling mechanism of superconductivity is governed primarily
by the spatial average of the interlayer one electron bare
hoping matrix element'.   We are of course inspired by Anderson's
theorem on dirty superconductor to call our semi quantitative
conclusion a theorem!

\section{Application to cuprates and organic superconductors}
It has been well established that the layer cuprate superconductors are
strongly correlated electron systems.  There are also
families of organic superconductors\cite{orgcond} which are strongly
correlated electronic systems.  Two major groups are the
TMTSF and the ET molecule (Bechgaard salt) based organic conductors.
The first group are
quasi one dimensional and the second one are quasi 2 dimensional.  These
are also essentially narrow one band systems in which electron-electron
interaction is important and  the scale of electron repulsion
energy is large compared to the one electron
band width.  Systematic experimental studies involving NMR,
photoemission and transport studies on the TMTSF based organic
conductor in its normal state has brought out the Luttinger
liquid character of the conducting chains.  The existence of
spin density wave states, spin Peierl's phase also brings out
the important of electron correlation in this one band tight
binding systems.
One of the remarkable manifestation of strong correlation,
is the recent suggestion that there is `confinement': Strong and
collaborators have suggested that the anomalous angular dependence of
the magneto resistance in  $(TMTSF)_2 X$  could be explained as a
commensurability  effect arising from confinement.

One of the puzzling and ubiquitous features of the organic
superconductors\cite{orgcond} is the
universal sensitivity of $T_c$ to off plane or off axis disorder.  It
is so dramatic that in
one of the ET based systems that in one case the $T_c$ is
reduced from 8 K to nearly 0 K by the disordering of the anion group,
which is non centro symmetric.  What is remarkable is that the disorder
is a physical effect rather than a chemical effect in the sense of not
changing the charge transfer to the planes and not changing the
nature of the chemical bonds.  The normal state properties
of the chains and planes are perhaps not strongly affected.

One of the systematically studied compound is the solid solution
$\beta(ET)_2 (I_3)_{1-x} (I Br_2)_x$.  The anions $I_3$ or $I
Br_2$ are located at crystallographic inversion centers.  Thus a
non centro symmetric anion like $I Br_2$ can be orientationally
ordered or disordered, even though positionally it is ordered.
For the case of $x =0$, the above compound  has a $T_c \approx 8 K$.
And for the case of $x = 1$, it has a $T_C
\approx 2K$.  The anions $I_3$ and $I Br_2$ continue to be
powerful acceptors and are in valence state of 1.  It is
found\cite{etds} that as $x$ is
varied continuously, $T_c$ falls rather fast and is essentially
zero in the range $ 0.2 < x < 0.7$.  In this region the anions
$I Br_2$ are orientationally disordered.

Similarly the quasi one dimensional conductor $(TMTSF)_2 Cl0_4$
exhibits singular suppression of $T_c$ with the anion disorder:
in this case also the non centro symmetric $Cl0_4$ ion is
situated at a crystallographic inversion center leading to a
possibility of the orientational disorder of the anions.

Even more remarkable is the salt $(ET)_2 I_3$ itself, which
can occur in three closely related crystallographic forms:
$\beta, \beta_c, \beta^*$, and their enomalous $T_c$
differences.  The ethylene side groups of the ET
molecule is capable of being in two conformations called the
eclipsed and staggered conformations.  In the above three
crystallographic forms the essential difference is in the
conformation of the ethylene side group.  If these two
conformations occur in a random fashion, the anion molecules are
also correspondingly disordered, leading to a large suppression
of $T_c$.

Some explanations\cite{fukuyama} invoking Anderson type of
localization induced in the
chains by the off axis randomness exists.  However, in view of the
reduction of Anderson localization phenomenon for weak disorder
in a strongly correlated system such as the organic conductors, one needs
a more satisfactory explanation.  Brazovskii and Yakavenko were
one of the first to study the sensitivity of $T_c$ to the type
of anion order in organic conductors.  Even though they talk about
conservation of coherence of cooper pairs between the conducting
chains, our mechanism that we will discuss here seems to be natural
and simple.

We argue that the off plane\/ chain randomness suppression the resonant
pair tunneling process is the major source of reduction of $T_c$.  The
non centro symmetric character of the anion group plays an important
role, as we demonstrate below.

The effective electronic Hamiltonian for these systems is a
spatially anisotropic Hubbard model:
\be
H = - \sum t_{ij} c^\dagger_{i\sigma} c_{j\sigma} + \mbox{h.c.}+
    U \sum n_{i\uparrow}n_{\downarrow}
\ee
The details of anisotropy is contained in the one electron
hoping matrix elements $t_{ij}$.  For the $TMTSF$ family of
conductors\cite{strong} the hoping matrix element along the chain is $\approx
0.25 eV$.  The hoping matrix element in the two directions normal
to the chain is low by about $\frac{1}{20}$ and $\frac{1}{30}$.
The on site U is at least twice as large as the band width along
the chain. For the ET salts the electronic parameters are
very similar except that it is quasi 2-dimensional.  There are
also strong electron phonon interactions (both intra molecular
and inter molecular), whose real role in stabilizing a
relatively high $T_c$ is not very clear.  It is likely, like in
the curates the effect strong electron correlation masks the
importance of electron phonon interaction.  Electron lattice
coupling stabilized their own phases such as Spin Peierls' with the
help of electron correlations for some value of the parameters.

For us the important thing to understand is the nature of the
bare one electron hoping matrix elements between two adjacent
molecules in the two planes or two chains.
In tight binding systems like the cuprates or the organic conductors, it
is natural to think, while discussing about one electron hoping matrix
element between adjacent layers, in terms of symmetry adapted molecular
or Wannier orbitals.  The tunneling is from one Wannier orbital of a
layer to a nearby Wannier orbital of the adjacent layer.  In view of the
large separation between the adjacent planes or chains, the direct
hoping matrix elements between the adjacent Wannier orbitals are
negligibly small.  Some LUMO bridging orbitals of the anion groups play
important role in establishing an appreciable bare one
electron  tunneling matrix element.

Let us take the case of TMTSF system . The relevant Wannier orbitals
of the two adjacent planes
are the HOMO of the TMTSF molecules $\phi_{i\alpha}, (\alpha = 1,2)$.
It
is a $p\pi$ bonded molecular orbital.  Direct tunneling matrix element
between them in
adjacent planes is negligibly small: $<\phi_{i1} | H | \phi_{i2}>
\approx 0$.  However, there is a finite overlap of the above orbital
with the LUMO of he $Cl O_4$ group.  The LUMO orbitals could be
degenerate in general.
Let us denote the bridging orbitals by $\phi_{b i\mu}$, where $\mu$ is
the degeneracy index and the subscript b stands for the bridging
orbital.  The effective one electron tunneling matrix element through
the bringing orbital in a second order process is given by
\be
t_{\perp i} \approx \sum_\mu \frac{<\phi_{i1} | H |
\phi_{bi\mu}><\phi_{b i\mu} | H | \phi_{i2}>}{E_i - E_b}
\ee
where $E_i$ and $E_b$ are the energy of the Wannier orbital of the
planes and the bridging orbital respectively.  And H is the one electron
Hamiltonian.

The LUMO's have in general nodes and change sign as we move within the
anion group.  Thus, when the anion molecules are disordered the matrix
elements $<\phi_{i1} | H | \phi_{b i \mu}>$ can change in sign and
magnitude depending on the
orientation of the anion molecule.  That is,
$<\phi_{i1} | H | \phi_{b i \mu}>$ can change in sign and
induce sign disorder in the bare one electron hoping matrix
element $t_{\perp i}$ (equation 18).

This is the way the disordered cation group can introduce disorder in
the sign of $ t_{\perp i}$,
This in turn suppresses the resonant pair tunneling by the
reduction in the spatial average value $<t_{\perp i}>$.  Using
our earlier
argument we have a simple prediction.  For a completely disordered
cation,
\be
<t_{\perp i}> = 0 \Rightarrow T_c \approx 0
\ee
In general if a fraction p of the anions are disordered,
\be
k_B T_c \approx \frac{(p - \frac{1}{2} )^2 <t_{\perp i}>^2}{4t}
\ee
We are able to make a reasonable fit of this with the
experimental results\cite{etds} on the solid solution $\beta (ET)_2 (I_3)_{1-x}
(I
Br_2)_x$ discussed
earlier.  It will be important to find this quadratic  decrease
of $T_c$ with the
degree of orientational disorder of the non-centro symmetric
anyons in organic superconductors in related systems.

we will now turn briefly to the case of cuprates, where the
nature of coupling between the planes is more complicated and has
different solid state chemistry.
It  has been experimentally seen\cite{jorgensen} that the oxygen vacancies in
the $CuO_2$ have little effect at low concentrations, while
those that form between the planes systematically lower $T_c$.
What is important is that it is not a mere charge transfer
change that reduces $T_c$.

It has been noticed\cite{jorgensen} that the structural
coherence of the $CuO_2$ planes, even though do not change the normal
state properties does affect the $T_c$ rather strongly.  It is generally
believed that both in the 214 and 123 compounds any tendency to have
orthorhombic short range order suppresses superconductivity.  A good
example could be
the 214 compound for the magic value of $x = 0.12$.  At this value $T_c$
almost vanishes.  Various explanations have been put forward to
understand this.  What is striking is that there is no true long range
orthorhombic order.  We believe  that the reduction in the pair
resonant pair tunneling matrix element caused by the structural
incoherence at atomic scales (owing to the well developed
short range orthorhombic distortions) in the $CuO_2$ planes is responsible
for the reduction in $T_c$.
The lack of structural
coherence in the plane and its environment severely suppression
the resonant tunneling of pairs between neighboring layers could
be an important source of $T_c$ reduction as we go to the
overdoped region, i.e. $x > .24$.  The conventional explanation
for the reduction in $T_c$ this region is overdoping and the
associated recovery of fermi liquid character.

Recently a family of $CuO_2$  layered superconductors also having
bridging $CO_3$ groups have been synthesised.  Unusual
sensitivity of
the superconducting $T_c$ to the disordering of the non centro symmetric
carbonate group has been observed often.  This could again be explained
by our mechanism of reduction of $T_c$ by c-axis disorder.

\section{Relation to Anderson's theorem on dirty
superconductors}
It is important to point out that Anderson's theorem\cite{pwads}
on dirty superconductor was proposed in the context phonon mediated
superconductivity.  We point out below that there are
important modifications to this theorem in the context of anisotropic
superconductors in the presence of strong correlations.
Soon after the appearance of BCS theory, Anderson explained the
unexpected  insensitivity of the superconducting $T_c$ on time
reversal symmetry respecting disorder like positional disorder
or alloying or non magnetic impurities.  Anderson's first point
with respect to setting up an effective BCS Hamiltonian was to
emphasize the formal use of the exact eigen states in the
presence of the random potential.  Electron phonon scattering
among these exact eigen states is used to construct the phonon
mediated two body interactions.  He emphasized the pairing among
time reversed eigen states as the basic and relevant two body
terms that will lead to a coherent superconducting state.

Our case is different and we can not use Anderson's theorem
directly.
In view of the phenomenon of confinement and in view of the fact
the system is anisotropic, to a first approximation the off
plane disorder does not affect the anomalous normal state.  This
means while employing a fermi liquid like BCS Hamiltonian for
the physics of interlayer pair tunneling a la' Anderson, one
needs to use only plane wave states.  The exact eigen states
of the one electron problem of the coupled planes (including
the random $t_{\perp i}$) is something that is not relevant
for the problem in the presence of strong correlation in the
plane.  In this sense the presence of strong correlation
defies Anderson's dirty superconductor theorem.

Recently Fay and Appel\cite{appel} have argued that it is the
non retarded character of superconductivity in cuprates that is
responsible for the violation of Anderson's theorem.  What we have
argued in our paper is a more fundamental reason which relies on the
resonant tunneling character of the interlayer pair
tunneling mechanism, and not on its non retarded character.

\section{Acknowledgements}
It is a pleasure to thank Prof T V Ramakrishnan and S V
Subrahmanyam for hospitality at the Indian Institute of
Science, Bangalore, where this work was presented in a
conference
on `High Temperature Superconductivity' in January 1993.
I also wish to thank the Indian PMB on Superconductivity and DST for
a grant.

\end{document}